# Growth of c-oriented MgB$_2$ thin films by Pulsed Laser Deposition: structural characterization and electronic anisotropy

C.Ferdeghini[1], V.Ferrando[1], G.Grassano[1], W.Ramadan[1], E.Bellingeri[1], V.Braccini[1], D.Marré[1], P.Manfrinetti[2], A.Palenzona[2], F. Borgatti[3], R. Felici[3], T.-L. Lee[4]

[1]INFM, Dipartimento di Fisica, Via Dodecaneso 33, 16146 Genova, Italy

[2]INFM, Dipartimento di Chimica e Chimica Industriale, Via Dodecaneso 31,16146 Genova, Italy

[3]INFM-Operative Group in Grenoble, c/o ESRF, BP 220, F-38043 Grenoble, France

[4]ESRF, BP 220, F-38043 Grenoble, France

**Abstract**

MgB$_2$ thin films were deposited using Pulsed Laser Deposition (PLD) and ex-situ annealing in Mg atmosphere. The films presented critical temperatures up to 36K and turned out to be preferentially c-oriented both on Al$_2$O$_3$ (r-cut) and MgO(100) substrates. Synchrotron analyses gave also some indications of in plane texturing. The films exhibit very fine grain size (1200Å in the basal plane and 100Å along c-axis) but the general resistivity behavior and the remarkable extension of the irreversible region confirm that the grains boundaries are not barriers for supercurrents. Upper critical field measurements with the magnetic field perpendicular and parallel with respect to the film surface evidenced a field anisotropy ratio of 1.8. The H$_{c2}$ values are considerably higher with respect to the bulk ones, namely when the field lies in the basal plane, and the field–temperature phase diagram for the two magnetic field orientations suggest the possibility of strongly enhancing the pinning region by means of texturing.

## 1.Introduction

The discovery of superconductivity with T$_c$≈40K in a simple binary structure as the MgB$_2$ one was quite a surprise [1]. Although this compound has been known and structurally characterized since the 1950's [2], no indications about its resistance al low temperature has been reported. Superconductivity in MgB$_2$ has aroused tremendous interest for the fundamental and practical aspects of this material. MgB$_2$ exhibits a lot of intriguing properties: it is the compound with the highest T$_C$ among non-oxide superconductors and initial reports indicate evidence of a phonon mediated mechanism for the superconductivity, making MgB$_2$ also the BCS material with the highest T$_C$ [3]. Grain boundaries have not dramatic effects on the critical current densities [4, 5, 6], being the coherence length of this compound longer than those of HTSC and thus avoiding the depression of the superconducting order parameter between grains. The above reported properties, the very simple structure and the commercial availability of this material make MgB$_2$ a favorite candidate for large scale and electronic applications and suggest the possibility of further progress in technology based on superconducting materials. The main limitation for large scale applications seems to be related to the considerable small value of the irreversibility field with respect to others technological superconductors as Nb-Ti [7].

Thin film in situ deposition is a quite difficult task; on the other hand, high quality epitaxial thin films are needed to implement a new class of electronic devices. Furthermore, due to the difficulty in the single crystal growth, the availability of epitaxial thin films is also important for research study. The main problems in thin film deposition are related to the high volatility of Mg, and to its high reactivity with oxygen. Different methods for superconducting MgB$_2$ film preparation have been reported in literature: all of them are based on a two-step process.

The first step is the deposition of a non superconducting precursor layer, which is then annealed in a second step to obtain a superconducting phase. The widely used precursor layers are amorphous Mg-B films [8-10], deposited at room temperature from a stoichiometric target, amorphous boron films [11, 12] or layers obtained from mixed B and Mg not stoichiometric target [13]. Commonly, annealing is performed ex-situ in magnesium atmosphere [8-12]; few authors present an in-situ, in vacuum treatments [10, 13]. Currently, samples with the highest T$_c$ are produced by ex situ annealing.

For electronic applications, wich require the integration of different materials as in devices or junctions, however, the annealing procedures present some drawbacks and an in situ deposition of superconducting MgB$_2$ is necessary: up to now, only one method to produce superconducting samples in one step has been reported in literature [14], but the samples have an onset critical temperature of only 28K and the process needs to be optimized.

In this paper, we report on the deposition of superconducting $MgB_2$ films by Pulsed Laser Deposition (PLD) via ex situ annealing process in magnesium atmosphere of a Mg-B precursor layer. We present also details on the optimization of the process, together with the structural and morphological characterization and some preliminary results on synchrotron measurements of the in plane texturing of the films.

Finally, the anisotropic electronic behavior of the upper critical field in c-oriented thin films, studied by means of transport measurements, is also described.

## 2. Pulsed laser ablation deposition of $MgB_2$ thin films

### 2.1. Deposition process

The PLD experimental apparatus consists of an UHV deposition chamber and a KrF excimer laser. Details of the apparatus are described in [15].

We used $MgB_2$ sintered target prepared by direct synthesis from the elements: Mg, in form of fine turnings (99.999 wt.% purity), and crystalline B, 325 mesh (99.7 wt.% purity), were well mixed together and closed by arc welding under pure argon into outgassed Ta crucibles which were then closed in quartz ampoules under vacuum. The samples were slowly heated up to 950°C and maintained at this temperature for 24 hours.

The target was then obtained by sintering these powders in pellets at 1100 °C for 72 hours following the same sealing procedure as above.

A transition temperature $T_C$ of 39K was determined from magnetization and resistivity measurements, and a residual resistivity ratio RRR=R(300K)/R(40K) of about 5 was determined using a four probe resistive method.

Different substrates were used in literature [8-13, 16, 17], namely $Al_2O_3$, MgO and $SrTiO_3$, with different crystallographic orientations. In general, substrates are chosen in such a way to have the same symmetry and a good lattice matching with the film. ($MgB_2$ has a hexagonal crystalline cell, with a=3.084Å and c=3.522 Å). Moreover, they must be compatible with the parameters used in the preparation process.

We chose two different substrates, MgO and $Al_2O_3$, because they are very stable at the high temperature used in the annealing process, so minimizing the film-substrate interaction.

Concerning MgO substrates we used, until now, the (100) crystallographic orientation which has a square surface symmetry with a=4.203Å. Nevertheless, it is used to drive the epitaxial growth of other intermetallic films, even if the cell parameters are very different [18]. More appealing appears the (111) crystallographic orientation, which presents a hexagonal surface symmetry, with a lattice mismatch with $MgB_2$ less than 3%.

Concerning sapphire, we used $Al_2O_3$ (r-cut); this substrate has been used by many authors for the growth of $MgB_2$ films, despite a not negligible mismatch of the lattice parameters and a not regular hexagonal surface cell.

Our precursor layers were deposited in vacuum, at room temperature. In fact it is well known that it is impossible to deposit $MgB_2$ thin films directly in at high deposition temperature, due to the high Mg volatility and also to the formation of Boron oxides, that have a low evaporation temperature [19]. To crystallize the superconducting phase, we carried out an ex-situ annealing procedure in magnesium vapor. The samples were placed in a tantalum box, which acts as an oxygen getter, containing Mg lump (approx 0.05 mg/cm$^3$), and then they are sealed in an evacuated quartz tube and heated at temperature ranging from 500°C to 900°C for 30 minutes followed by rapid quenching to room temperature. The final superconducting properties strongly depend on the annealing temperature.

In figure 1 the dependence of the transition temperature on the annealing temperature is shown: the superconducting phase begins to form at temperature about 550°C, and the transition temperature increases monotonically with the annealing temperature, approaching 36K at 850°C. The samples annealed at higher temperature are semiconducting and they do not show superconducting transition.

The superconducting samples annealed at lower temperatures show a slightly semiconducting behavior in the normal state resistivity, while the samples presenting $T_c$ higher than about 30K present a slightly metallic behavior, with a maximum RRR of about 1.5.

## 2.2. Structural analysis by means of standard x-rays diffractrometry and synchrotron radiation

To evaluate the structural properties of the samples, we performed standard x-rays θ-2θ analysis in Bragg-Brentano geometry. The diffraction patterns show the presence of mainly (002) $MgB_2$ reflection with small peaks coming from different orientations and from other phases. Actually $MgB_2$ peaks were detected only in samples that exhibit $T_c$ higher than ~30K. The predominance of (002) reflection, with respect to the other $MgB_2$ reflections, indicates that our films are preferentially c-oriented.

In figure 2 the rocking curve around the (002) peak of $MgB_2$ is reported. The curve has a full width at half maximum (FWHM) value of about 5°, confirming the good orientation of our samples. In samples deposited on $Al_2O_3$, the (00l) $MgB_2$ reflections are hidden by the very intense substrate peaks.

Because of the low scattering factor of Mg and B, the diffraction peaks have very low intensity, and we chose to perform also detailed analyses by means of high brilliance synchrotron radiation.

Structural characterization of $MgB_2$ films has been carried out using the ID32 beamline of ESRF. The ID32 diffractometer is a 2+2 geometry [20] with the sample mounted vertically. The used incident photon energy was 12.45 keV produced by two undulators and monochromatized by a double Si(111) monochromator. The sample was mounted in order to have its surface normal parallel to the azimuthal rotation angle. The crystallographic alignment has been carried out taking as reference reflections two peaks of the sapphire substrate.

Two distinct analysis have been carried out. First, we have measured the diffracted intensity along a direction close to the surface normal. This is analogous to a standard ϑ-2ϑ scan where the diffracted intensity has been collected as a function of the incidence angle maintaining the exit angle equal to the incident value minus 1° to avoid the strong substrate peaks. The results of these kind of scans are shown in figure3 as a function of the 2ϑ angle where the substrate peaks have been subtracted. Several peaks belonging to the film are clearly observable. In particular the three peaks at 2ϑ 16.4° and 33.2° contain the signal coming from the (001) and (002) reflections of the stechiometric $MgB_2$ phase.

Because their position is practically coincident with the substrate sapphire peaks, they can only be observed after having misaligned the substrate orientation from the angle o incidence equal to the angle of exit condition. The integrated intensities of these peaks scale very well with the tabulated values, in fact we measure a ratio of I(001)/I(002) of 0.23 with respect to an expectation value of 0.27. Considering that we have not corrected these values for the absorption, because of the uncertainties in the sample thickness, and for the geometrical corrections which would cause an increase in the measured ratio we judge the agreement more than satisfactory. The other peaks at 24.5° and 28.3° are due to different phases of magnesium-boron compounds present in the film. The small peak at 27.2° is probably due to the (101) $MgB_2$ reflection indicating that a small fraction of this phase could be disordered. We have to consider that the tabulated intensity of the (101) reflection should be about 8 times bigger than the (002) reflection while we observe a ratio of 1/50.

To determine whether the $MgB_2$ film is also oriented in plane we have carried out measurements with incidence and exit angles of 0.5° as a function of a vertical $2\vartheta$ angle (this implies that the scattering vector practically parallel to the sample surface) and of the azimuthal sample angle. Several peaks have been observed and among them we were able to observe one which is due to the (110) $MgB_2$ reflection. This peak has a strong azimuthal dependence indicating that the film has an in plane preferential orientation with the main axis along the substrate [110] direction.

**2.3. Morphologic characterization**

In order to evaluate the effect of the annealing procedure on the sample morphology, AFM (Atomic Force Microscopy) measurements were performed before and after the thermal treatment. Figure 4 (image on the left) shows the surface of a precursor film having a RMS roughness of about 15 nm. Similar values were measured after the annealing ( image on the right side of figure 4). This result was obtained for the whole annealing temperature range, indicating that the annealing procedure does not strongly affect the surface morphology.

The in plane dimension of the grains, as measured from AFM images, is about 120nm, and it does not change after and before the annealing treatment. The grain size values are comparable with the values reported in [16] and [17].

From the broadening of the (002) reflection in the θ-2θ scans, we estimated the grain size along the growth direction. By using the Sherrer formula d=0.9 λ/ FWHM cosθ, where λ is the x-rays wavelength we obtained a value of about 10nm. Therefore, the films resulted to be nanostructured both in plane and along the c-axis

**3.Electronic anisotropy**

Due to the layered structure of $MgB_2$ compound, an electronic anisotropy should be expected. Precise anisotropy determination requires quite large single crystals, not immediately available. An approximate calculation was performed [24] by using aligned crystallites: the anisotropy factor η (i.e. the ratio between the critical fields parallel and perpendicular to the planes) resulted to be 1.7, as estimated from ac susceptibility measurements. Measurements of the anisotropy ratio performed on powder samples by means of conduction electron spin resonance gave a value of 6-9 as a result. [21]. Recently measurements on c-oriented thin films appeared in literature. Patnaik et al. [22] reported anisotropy measurements on films grown on $SrTiO_3(111)$: they found, in three films with different normal state resistivity ranging from 38 to 360μΩcm, an anisotropy factor η in the range 1.8-2. M.H.Jung et al [23], instead, found a lower value η=1.25, despite they claim for epitaxiality of their film grown on $Al_2O_3(102)$ with very low normal state resistivity (5μΩcm). The very recent availability of small size single crystal does not definitively clarify the topic. In fact, the experiments performed on single crystals gave values in the range 2.6-3 [25-27]. Therefore the actual η value needs to be confirmed.

As shown in the previous section, our films are preferentially c-oriented both on sapphire and MgO substrates: we therefore performed electrical resistivity measurements with magnetic field applied perpendicular and parallel to the *ab* planes.

Measurements were performed on the film deposited on MgO(100), whose rocking curve around the (002) reflection is reported in fig.2. The sample exhibits $T_C$=31.4K with $\Delta T_C$=1.1K and a RRR=1.16 and a normal state resistivity, just above the transition, $\rho_N \approx$

100μΩcm. The $\rho_N$ value must be considered with an error of 20% due to the uncertainty on the estimated thickness, done by means of x rays reflectivity measurements. Electrical resistance measurements as a function of temperature in applied magnetic field up to 9T were performed in a Quantum Design PPMS apparatus by using a four-probe AC resistance technique at 7 Hz and a feeding current of 0.5mA. The current density was perpendicular to the magnetic field. Electrical contacts were made with aluminum wires bonded to the sample by an ultrasonic bonding machine.

Figure 5 shows the resistive transitions with the fields applied perpendicular (upper panel) and parallel (lower panel) to the film surface.

The anisotropic behavior is clearly evident, being the critical temperature strongly depressed by the magnetic field applied in the perpendicular direction, and the transition width for H parallel is narrower than that of H perpendicular (transition widths at 9T are 1.8K and 3.6K for the parallel and perpendicular configuration respectively).

$H_{c2}$ and $H_{irr}$ were estimated from resistivity measurements. $H_{c2}$ vs. temperature curves were determined as the midpoint of the resistive transition for each field while the temperature at which the resistance vanishes gives the irreversibility line. In figure 6 we report $H_{c2}$ and $H_{irr}$ as a function of temperature for both orientations: $H_{c2}$ and $H_{irr}$ are considerably higher when the field is parallel to the film surface.

From these data it is possible to calculate the temperature independent anisotropy factor η that resulted to be 1.8. This value is in a good agreement with [22]. In any case this value must be considered only as a lower limit due to the not complete c-orientation of the sample. In fact, also the (110) reflection in the X-rays pattern was detected, even if with low intensity. In the same figure, as a comparison, $H_{c2}$ and $H_{irr}$ of a sintered sample are also reported (the same bulk that we used as target).

We can observe, for the film, a linear $H_{c2}$ versus T dependence near $T_C$: in fact, none of the two configurations shows the upward curvature typical for the bulk samples (see the data for the bulk in figure 6). This curvature is related to the clean limit condition. The linear dependence found in the case of the film can be accounted for the disorder and impurities introduced during the growth. Both reduce not only the critical temperature, but also the electronic mean free path therefore inducing the dirty limit condition. Our film presents a normal state resistivity of about 100µΩcm: this high value suggests that the dirty limit condition is reached.

To verify this fact, from the normal state resistivity we can estimate the mean free path using the free electron mass, the reported Hall coefficient [28] and the Fermi velocity from band calculation [29]. The mean free path is estimated to be of the order of 1 Å. This unphysical value, coming from the high $\rho_0$ value, can be explained in term of the very small grain size in the film. In fact, this can affect the effective section but also can introduce a grain boundary resistivity contribution that is supposed to be the cause for the lack of the anomaly at Tc in the thermal conductivity measurements [30]. As stated before our film shows a normal state resistivity value of about 100µΩcm, 20 times higher in respect to the film in ref [23] but intermediate in the range 38-360µΩcm reported in ref [22]. Patniak et al found a direct $H_{c2}$ dependence on the normal state resistivity values and an inverse $T_c$ dependence. Our film is in trend with respect to these data and shows a similar very fine grain size. Therefore we hypothesize a conspicuous geometrical effect in resistivity measurements.

However, by using the dirty limit extrapolation

$$H_{c2}(0) = 0.69 \frac{dH_{c2}(T)}{dT} T_C$$

high values of $H_{c2}(0)$ are found.

In particular for the two configurations we have $H_{c2}^{par}$ = 26.4T and $H_{c2}^{perp}$ =14.6T. These values are considerable higher with respect to the bulk ones.

Another interesting feature of the data in figure 6 is related to the irreversibility line behavior. It is clear that the separation between $H_{c2}$ and $H_{irr}$ is larger in the case of the bulk sample. While $H_{irr}(T) \approx 0.6 H_{c2}(T)$ in the case of the bulk, for the film we found: $H_{irr}^{perp}(T) \approx 0.8 H_{c2}^{perp}(T)$. It must be noted that this good pinning properties of our film with respect to sintered samples are present despite the nanometric size of grains; this supports the hypothesis that boundaries can act as pinning centers.

From these data the advantage in texturing $MgB_2$ is clear. In fact, although the current is not limited by the grain boundaries, the lower $H_{irr}$ value limits the current transport capability.

In conclusion, we have deposited MgB$_2$ thin films by Pulsed Laser Deposition. The best films present T$_c$ as high as 36K and are mainly c-oriented, both on MgO and sapphire substrates, and preliminary synchrotron measurements presented evidences of in plane texturing of the films. Electronic anisotropy measurements indicate an upper critical field anisotropy ratio of 1.8 and, moreover, the boundaries of the fine superconducting grains do not strongly affect the supercurrents of the films, with respect to bulk samples, and the enhanced field-temperature phase diagram of the films suggests the possibility to use the texturing as a way to improve the pinning properties of the samples.

**Acknowledgments**

We are pleased to thank J. Zegenhagen for having given us the possibilities of accessing the ID32 beamline.

**Figure caption**

Fig.1 Film critical temperature as a function of annealing temperature.

Fig.2 Rocking curve around the (002) reflection for film grown on MgO(001).

Fig. 3 $\vartheta$-$2\vartheta$ intensity scan measured from the MgB$_2$ film measured with a misalignment of 1° with respect to the surface normal after having subtracted the contribution from the substrate. The peaks at 16.4° and 33.2° are the (001) and (002) MgB$_2$ peaks. The other observed peaks at 24.5° and 28.33° are due to non stechiometric phases of the compound. The small peak at 27.2° a position could be due the (101) reflection of the MgB$_2$ indicating that part of the film is disordered.

Fig.4 AFM images performed on the precursor layer (on the left) and on film annealed at 750°C (on the right).

Fig.5 Resistive transitions with different magnetic fields, applied both perpendicular (upper panel) and parallel (lower panel) to the film surface. The anisotropy is evident.

Fig.6 H$_{c2}$ (open symbols) and H$_{irr}$ (full symbols) for the two magnetic field configurations for the c-oriented film. The curves for a sintered bulk are shown in the graph as a comparison. For this purpose the data are reported as a function of reduced temperature.

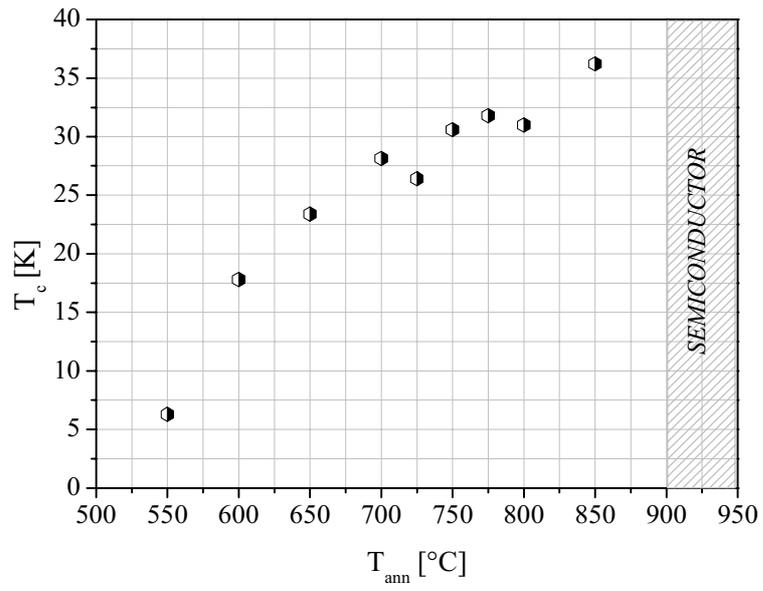

Figure1

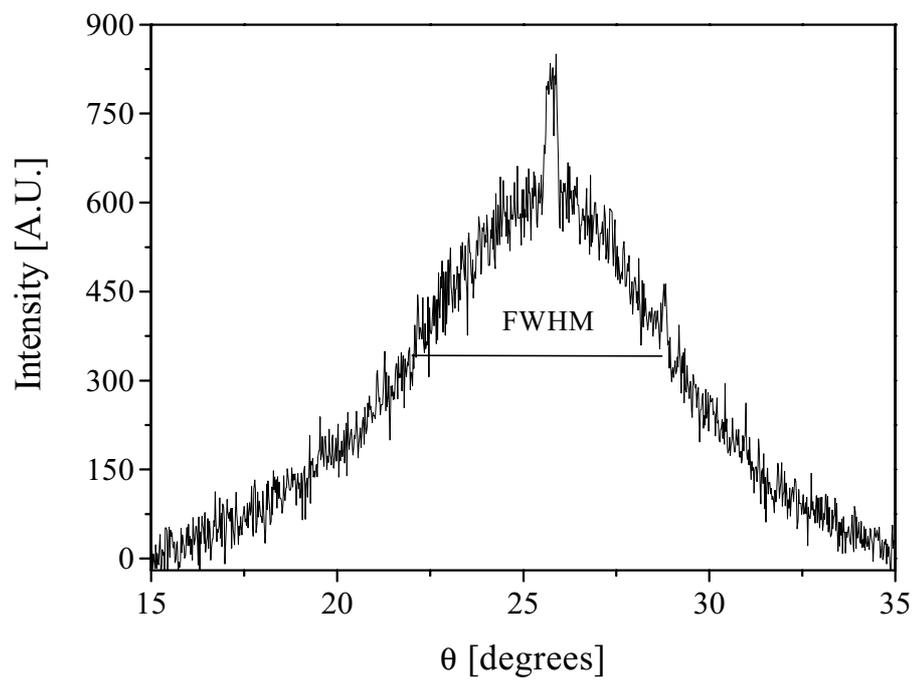

Figure 2

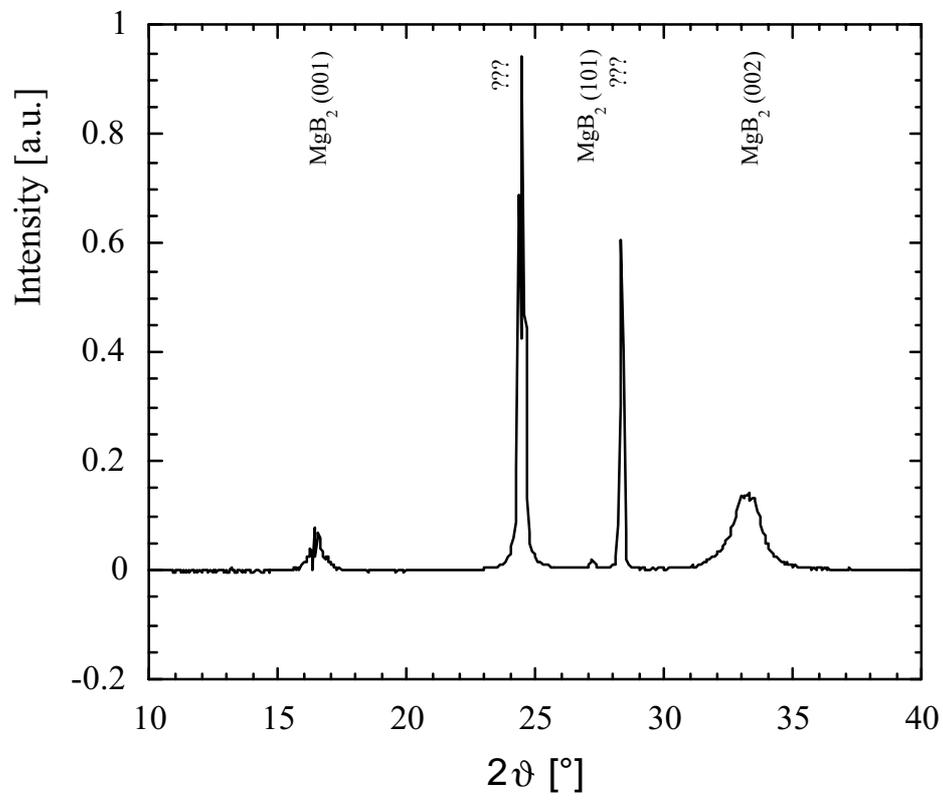

Figure3

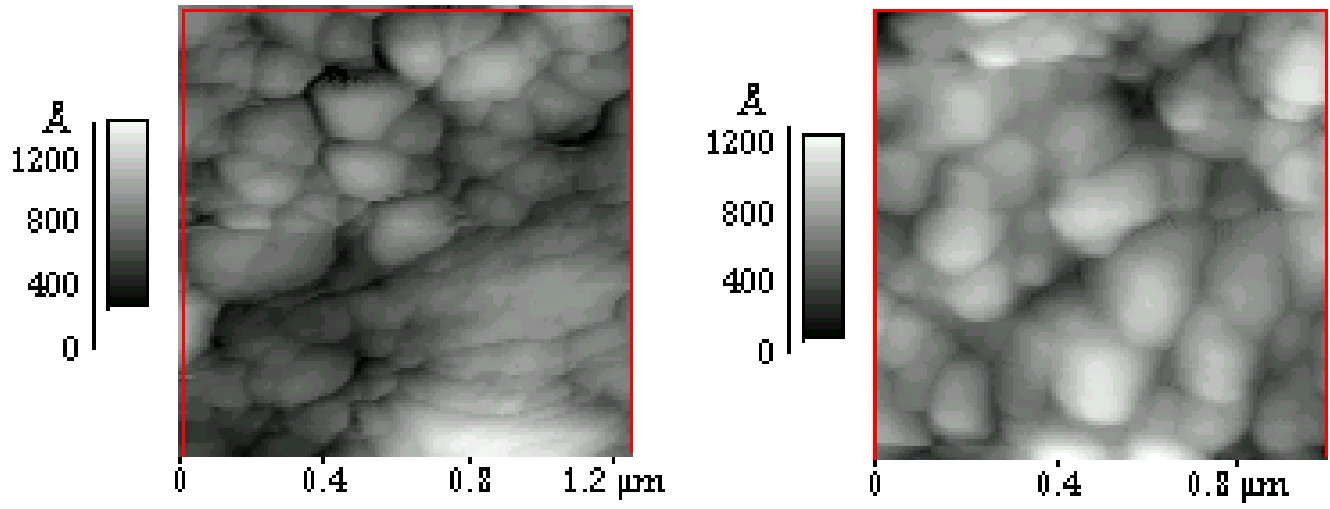

Figure 4

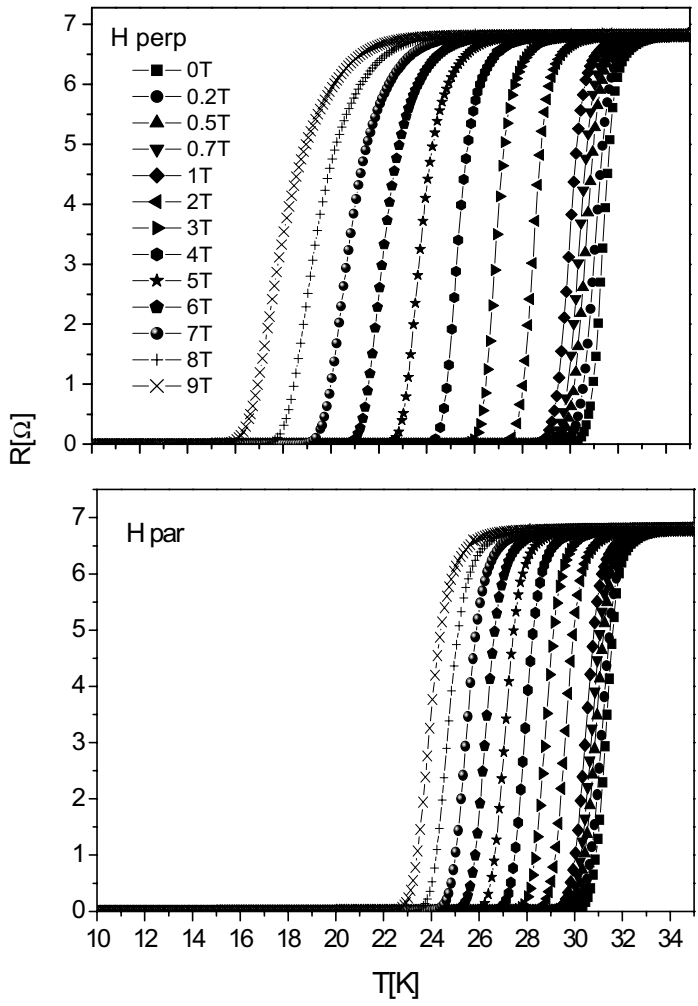

Figure5

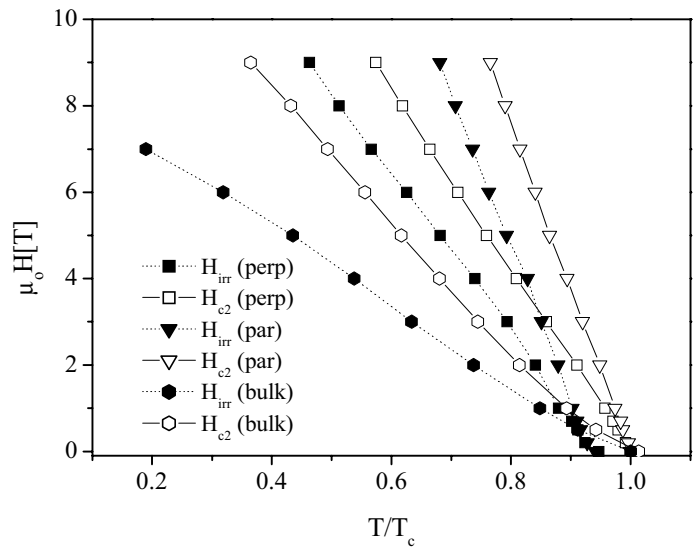

Figure 6